\documentclass[twoside,journey]{IEEEtran}
\usepackage{makecell}
\usepackage{array}
\usepackage{graphicx,amssymb,amsmath}
\usepackage{multicol}
\usepackage[noadjust]{cite} 
\usepackage{setspace}
\usepackage{subfigure}
\usepackage{graphicx}
\usepackage{float}
\usepackage {url}
\usepackage{stfloats}
\usepackage{amsthm,pifont}
\usepackage{flushend}
\usepackage{cases,subeqnarray}
\usepackage{bm,multirow,bigstrut}
\usepackage{amsmath, amsthm, amssymb}
\usepackage{textcomp}
\usepackage{latexsym,bm}
\usepackage{booktabs}
\usepackage{xcolor}
\usepackage{mathtools}
\usepackage{dsfont}
\usepackage{extarrows}
\usepackage{epsfig}
\usepackage{epsfig}
\usepackage{epstopdf}
\usepackage[noend]{algpseudocode}
\usepackage{algorithmicx,algorithm}

\theoremstyle{plain}

\theoremstyle{plain}

\usepackage{amsmath}

\IEEEoverridecommandlockouts
\begin{document}
\title{Extremely Large-Scale MIMO: Fundamentals, Challenges, Solutions, and Future Directions
\thanks{Zhe Wang, Jiayi Zhang, and Bo Ai are with Beijing Jiaotong University; Hongyang Du and Dusit Niyato are with Nanyang Technological University; Wei E. I. Sha is with Zhejiang University; M{\'e}rouane Debbah is with Technology Innovation Institute and also with University Paris-Saclay.}
}
\author{Zhe Wang, Jiayi Zhang,~\IEEEmembership{Senior Member, IEEE}, Hongyang Du, Wei E. I. Sha,~\IEEEmembership{Senior Member, IEEE},\\ Bo Ai,~\IEEEmembership{Fellow,~IEEE}, Dusit Niyato,~\IEEEmembership{Fellow,~IEEE}, and M{\'e}rouane~Debbah,~\IEEEmembership{Fellow,~IEEE}}
\maketitle
\vspace{-1cm}
\begin{abstract}
Extremely large-scale multiple-input-multiple-output (XL-MIMO) is a promising technology to empower the next-generation communications. However, XL-MIMO, which is still in its early stage of research, has been designed with a variety of hardware and performance analysis schemes. To illustrate the differences and similarities among these schemes, we comprehensively review existing XL-MIMO hardware designs and characteristics in this article. Then, we thoroughly discuss the research status of XL-MIMO from ``channel modeling'', ``performance analysis'', and ``signal processing''. Several existing challenges are introduced and respective solutions are provided. We then propose two case studies for the hybrid propagation channel modeling and the effective degrees of freedom (EDoF) computations for practical scenarios. Using our proposed solutions, we perform numerical results to investigate the EDoF performance for the scenarios with unparallel XL-MIMO surfaces and multiple user equipment, respectively. Finally, we discuss several future research directions.
\end{abstract}
\begin{IEEEkeywords}
Future wireless communications, XL-MIMO, near-field communications, degree-of-freedoms.
\end{IEEEkeywords}
\IEEEpeerreviewmaketitle
\section{Introduction}
\begin{figure}[t]
\setlength{\abovecaptionskip}{-0.1cm}
\centering
\includegraphics[width = 0.46\textwidth]{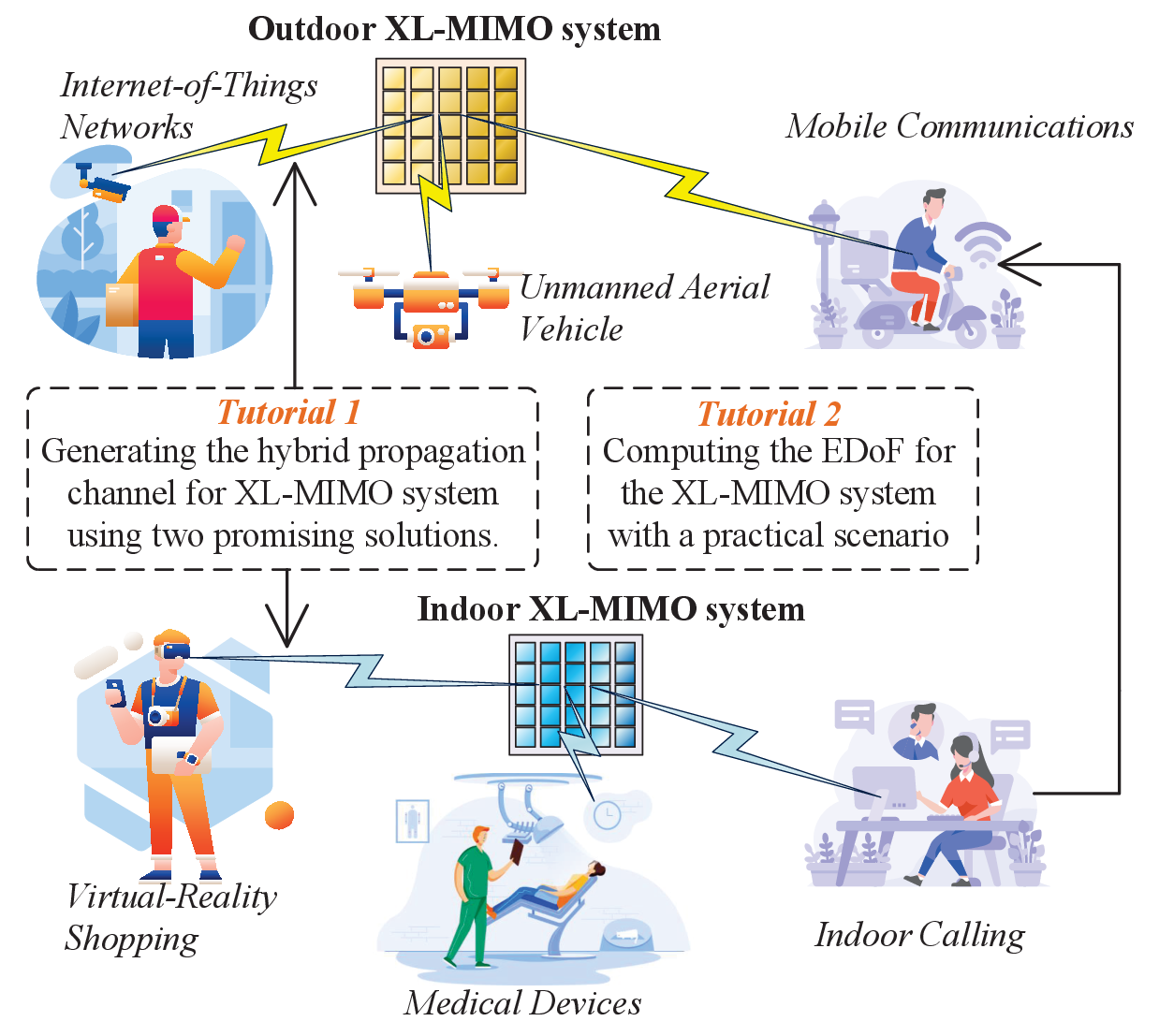}
\caption{Application scenarios for the XL-MIMO system.}
\label{fig1:four_models1}
\end{figure}

Next-generation wireless communication networks, such as beyond fifth generation (B5G) and sixth generation (6G) networks, are expected to meet the increasingly demanding Internet service requirements, e.g., ultra-low access latency, high transmission data rate, and extremely low bit error probability. Among numerous 6G technologies, the extremely large-scale multiple-input-multiple-output (XL-MIMO) technique takes a place through its ability to provide high spectral efficiency (SE), high energy efficiency (EE), and reliable massive access. The main application scenarios for the XL-MIMO are depicted in Fig. \ref{fig1:four_models1}.

Compared with the conventional massive MIMO (mMIMO), the basic idea of the XL-MIMO is to deploy an extremely large (perhaps infinite) number of antennas in a compact space. Specifically, many XL-MIMO hardware design schemes with different structures and terminologies have been investigated, e.g., \emph{holographic MIMO} \cite{9716880,9779586,9848831}, \emph{large intelligent surfaces (LIS)} \cite{8319526,9139337}, \emph{extremely large antenna array (ELAA)} \cite{2022arXiv220503615L}, and \emph{continuous aperture MIMO (CAP-MIMO)} \cite{2021arXiv211108630Z}. All these schemes are regarded as promising solutions for future wireless communications. However, their characteristics and relationships have not been well discussed.

In addition to the difference in hardware structure, the XL-MIMO introduces several new features that the conventional mMIMO does not have. The reason is that the XL-MIMO not only increases the number of antennas but also causes a fundamental change in electromagnetic (EM) characteristics to improve significantly the spectral efficiency and spatial degrees of freedom (DoF). However, the new features of XL-MIMO make system performance analysis and optimization difficult, because the XL-MIMO pushes the EM operating region from the far-field region to the near-field one.
In the conventional mMIMO system, the boundary to divide the near-field region and the far-field region, i.e., the Rayleigh distance, is negligible. Therefore, we can assume that the receiver is located in the far-field region and the EM wave can be simply modeled based on the planar wave assumption, where all array elements have the same signal amplitude and angle of arrival/departure (AoA/AoD). However, the far-field assumption fails in the analysis of the XL-MIMO system, because the significant increase in the number of antennas and antenna apertures cannot make the Rayleigh distance negligible~\cite{2022arXiv220503615L,9314267}. The receiver is very likely to be located in the near-field region, where the EM wave should be described based on the vectorial spherical wave assumption. We should also consider that the distances and AoA/AoD between the transmitter and receiver antenna elements vary over the antenna array. Besides, the channel spatial non-stationarity should be investigated, which is generally not considered in the studies of conventional mMIMO. When addressing the channel spatial non-stationarity, only a portion of the array is viewable to the users, so-called the Visibility Region. Many efforts have been endeavored to derive the accurate and tractable channel modeling methods for the line-of-sight (LoS) propagation \cite{9650519,9617121} or non-line-of-sight (NLoS) propagation \cite{2022arXiv220503615L,9765526} in the XL-MIMO system. However, the aforementioned difficulties in the near-field region analysis lead to insufficient discussion in a practical XL-MIMO channel model that considers both the LoS and NLoS propagation components simultaneously.

Moreover, to measure the performance of the XL-MIMO systems, significant performance metrics such as DoF~\cite{9139337} and effective DoF (EDoF)~\cite{9650519} are investigated, by using the electromagnetic information theory (EIT)~\cite{2021arXiv211108630Z}. Although the superior performance of the XL-MIMO has been reported, current research efforts in DoF analysis are based on primitive scenarios such as parallel transmitter and receiver, single transmitter and receiver pair, and scalar Green’s function-based channel model. It is necessary to explore the DoF performance in more practical XL-MIMO systems. Meanwhile, to capture the actual performance, several works on signal processing of the XL-MIMO system have also been implemented, such as the reduced-subspace-based channel estimation scheme \cite{9716880}, the polar domain-based channel estimation scheme \cite{2022arXiv220503615L} and the pattern-division multiplexing technique \cite{2021arXiv211108630Z}. Moreover, when the XL-MIMO system is implemented for the broadband communications, the near-field beam squint should be considered. The existing schemes addressing the far-field beam squint would no longer perform well in the XL-MIMO system. However, due to the large array size for the XL-MIMO system, processing schemes with acceptable complexity should still be explored.

Compared with the conventional mMIMO, the XL-MIMO mainly shows the following characteristics: 1) \textbf{\emph{Larger number of antennas}}: The XL-MIMO has higher order-of-magnitude in the number of antennas compared with mMIMO. Thus, the hardware design of XL-MIMO is different from that of mMIMO. 2) \textbf{\emph{New channel characteristics}}: The channel for the XL-MIMO should be modeled based on the vectorial spherical wave due to the near-field EW characteristics, while the channel for the mMIMO is modeled based on the plane wave with the far-field assumption. Besides, the channel spatial non-stationarity, mutual coupling, and polarization should also be considered in the XL-MIMO, which are omitted in the mMIMO. 3) \textbf{\emph{Near-field based signal processing schemes}}: Due to the different channel characteristics with the mMIMO, the XL-MIMO should be implemented with brand-new near-field based processing schemes, e.g. the channel estimation schemes based on the near-field channels and the near-field beam focusing.

In this article, we investigate the XL-MIMO technique for future wireless communications. The main contributions are summarized as follows:
\begin{itemize}
\item We present seminal XL-MIMO schemes and summarize general XL-MIMO hardware design schemes. More importantly, we comprehensively introduce their characteristics and clearly illustrate their relationships.
\item The fundamentals of the XL-MIMO are reviewed thoroughly from the aspects of ``channel modeling", ``performance analysis", and ``signal processing". For instance, Green's function-based channel modeling for the LoS propagation and the Fourier plane-wave representation for NLoS propagation is introduced. The readers can capture the comprehensive channel modeling methods for the XL-MIMO.
\item We discuss some existing challenges for the XL-MIMO and propose promising solutions. More importantly, we provide two use cases for the modeling of hybrid propagation channels and the EDoF computations for practical scenarios. And numerical results are given to investigate the EDoF performance for the scenarios with unparallel XL-MIMO surfaces or multiple user equipments. Based on our investigated solutions, the readers can implement further research on the channel modeling and performance optimization. Finally, some research directions are highlighted for further investigation.
\end{itemize}

\section{Fundamental Theories of XL-MIMO}
In this section, we introduce existing representative XL-MIMO hardware design schemes and show their relationships. Then, we provide a detailed review of the fundamentals of XL-MIMO from three perspectives, i.e., ``channel modeling", ``performance analysis" and ``signal processing".

\subsection{General XL-MIMO Hardware Design Schemes}
\begin{figure*}[t]
\setlength{\abovecaptionskip}{-0.1cm}
\centering
\includegraphics[width = 1\textwidth]{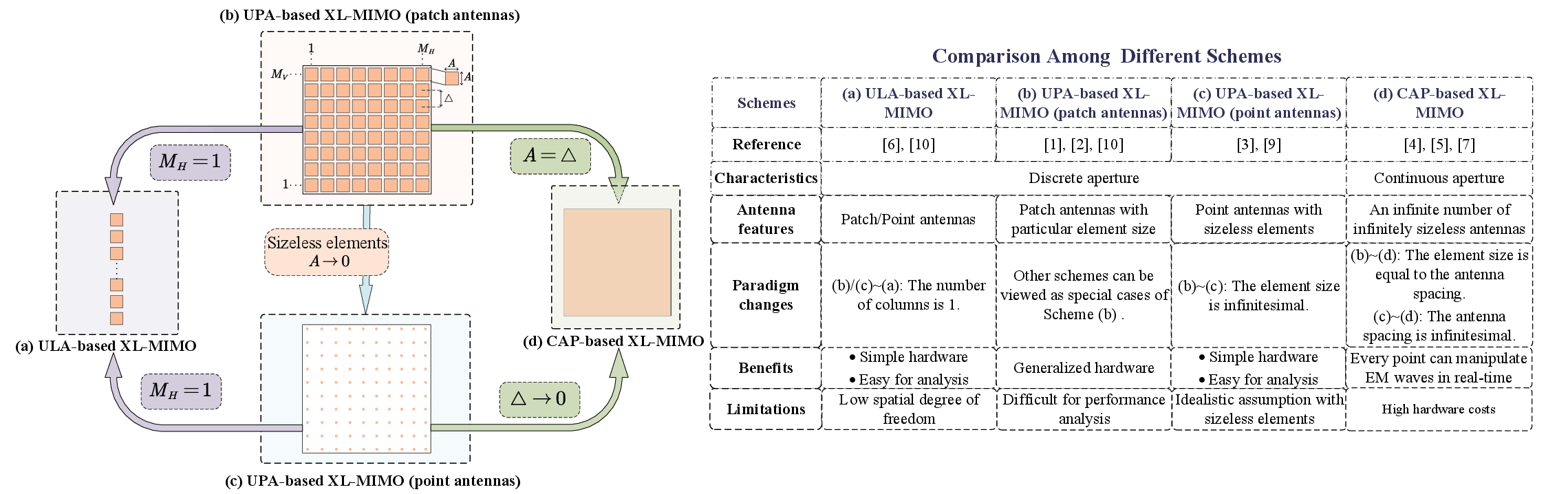}
\caption{General XL-MIMO schemes and their characteristics.$\bigtriangleup$ is the horizontal and vertical antenna spacing. $A$ is the element size. $M_V$ and $M_H$ are the numbers of antennas per row and per column, respectively. Besides, the isotropic patch or point antennas are assumed in this paper. Note that the CAP-based XL-MIMO with spatially-continuous aperture and the other schemes with discrete antennas are different on the practical design and the signal processing protocol. However, to comprehensively embrace the XL-MIMO schemes, we also discuss the promising CAP-based XL-MIMO and give some intuitive mathematical and physical relationships between the CAP-based XL-MIMO and the other XL-MIMO schemes.} 
\label{fig1:four_models}
\end{figure*}

As a promising technology for next-generation wireless communication, the XL-MIMO has been widely investigated and many XL-MIMO hardware design schemes have been proposed. To clearly show the characteristics of these existing schemes and how they are physically different, Fig. \ref{fig1:four_models} illustrates four general XL-MIMO hardware design schematics and describes their characteristics and relationships.

\textbf{\emph{(a). Uniform Linear Array (ULA)--based XL-MIMO}}: The first widely studied scheme is to equip the base station (BS) with an extremely large-scale ULA \cite{2022arXiv220503615L}, as shown in Fig. \ref{fig1:four_models} (a). Compared with the current massive MIMO system (64 or 128 antennas only), the extremely large-scale uniform linear array is expected to have another order-of-magnitude antenna number, e.g. 1024 antennas or more. Compared with the conventional massive MIMO, having another order-of-magnitude antenna number means not only an increase of the number of antennas but a change in the EM characteristics.

\textbf{\emph{(b)/(c). Uniform Planar Array (UPA)-based XL-MIMO with patch/point antennas}}: For these schemes (Figs. \ref{fig1:four_models} (b) and (c)), the BS is equipped with a UPA with a large number of antennas, e.g. thousands of antennas. The horizontal or vertical antenna spacing is usually below half of the wavelength, where densely packed sub-wavelength patch antennas with a particular size are incorporated \cite{9716880,9617121}. Moreover, one convenient implementation for the UPA-based XL-MIMO is to model the transmitting antennas at the BS as point sources, which is widely assumed for its convenience of analysis. More specifically, we can consider this model by regarding the patch antennas in the UPA-based XL-MIMO with patch antennas as sizeless point antennas \cite{9650519}.

\textbf{\emph{(d). Continuous Aperture (CAP)-based XL-MIMO}}: To improve the communication performance with limited apertures, continuous aperture (CAP)-MIMO \cite{2021arXiv211108630Z}, also known as holographic MIMO surfaces or large intelligent surfaces (LIS) \cite{9139337} (Fig. \ref{fig1:four_models} (d)), is proposed. The CAP-MIMO is composed of an infinite number of infinitely sizeless antennas. Different from the above three XL-MIMO schemes with a large number of discrete antennas over particular antenna spacing, the CAP-MIMO adopts the form of an infinite number of antennas with infinitely small antenna spacing and therefore can be viewed as a spatially-continuous EM surface.

We notice that three other schemes can be viewed as special cases of the UPA-based XL-MIMO with patch antennas.

$\bullet$ \emph{UPA-based XL-MIMO (patch antennas)} $\rightarrow$ \emph{UPA-based XL-MIMO (point antennas)}

The patch antenna-based surface can be considered as a special case with the sizeless point antenna-based surface by adjusting the element size to be close to $0$. This UPA-based XL-MIMO with sizeless point antennas is a reasonable assumption of the UPA-based XL-MIMO with patch antennas and is convenient for the channel estimation and performance analysis.

$\bullet$ \emph{UPA-based XL-MIMO (patch/point antennas)} $\rightarrow$ \emph{ULA-based XL-MIMO}

We can consider the ULA-based XL-MIMO as a special case of the UPA-based XL-MIMO with patch/point antennas with only one column of patch/point (other shapes would be suitable) antennas.

$\bullet$ \emph{UPA-based XL-MIMO (patch/point antennas)} $\rightarrow$ \emph{CAP-based XL-MIMO}
When the element size of the patch antenna is equal to the antenna spacing, the discrete UPA with patch antennas becomes the continuous-aperture surface. In addition, when the antenna spacing is infinitesimal, the UPA with point antennas can also become the continuous-aperture surface.

\subsection{Channel Modeling}\label{Sec_Channel}
\begin{figure*}[t]
\setlength{\abovecaptionskip}{-0.1cm}
\centering
\includegraphics[width = 0.82\textwidth]{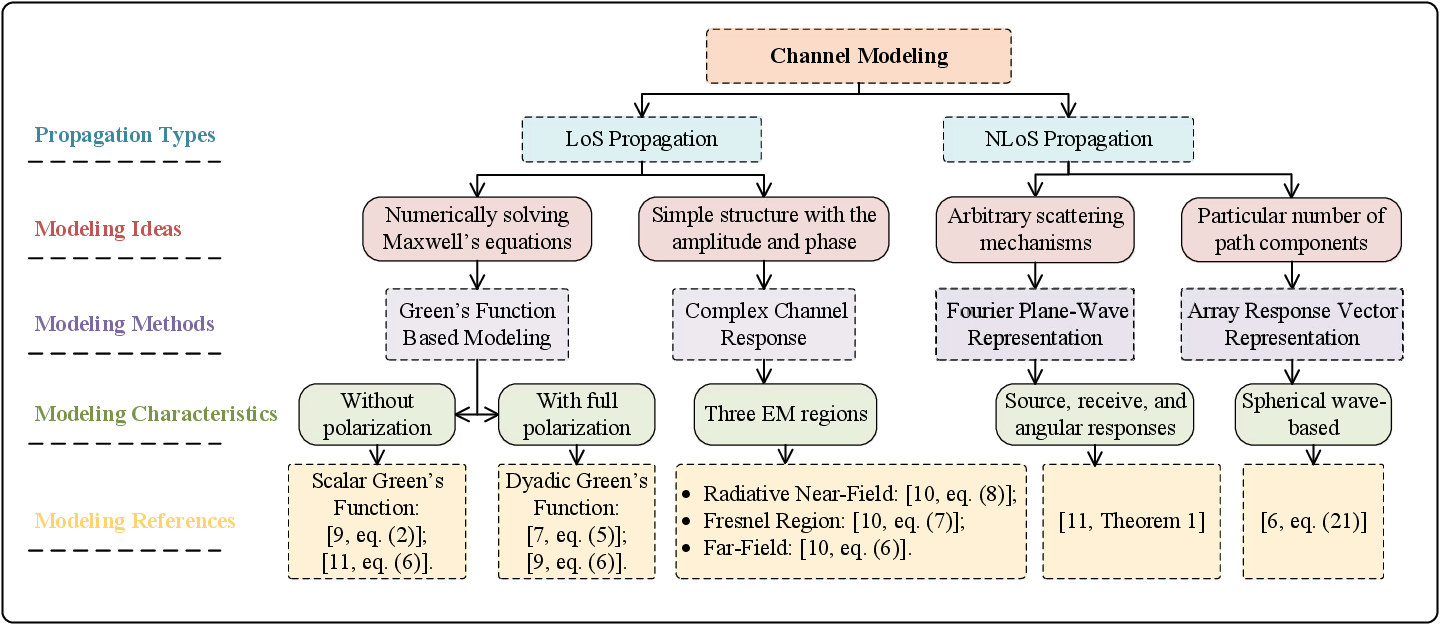}
\caption{Fundamentals of the XL-MIMO channel modeling.}
\label{Channel}
\end{figure*}

In the XL-MIMO, the receiver is very likely to be located in the near-field region, where the EM wave should be accurately described based on the vectorial spherical wave assumption. More importantly, any electric current density distribution can be generated at the transmitter to flexibly design the radiated EM field, and corresponding incident electric fields are induced at the receiver through the free-space propagation environment (LoS) or the arbitrary scattered propagation environment (NLoS). Based on the above descriptions, we introduce the fundamentals of channel modeling for LoS/NLoS propagation, which have been summarized in Fig. \ref{Channel}.

\subsubsection{LoS Propagation}
The XL-MIMO with a large array size shortens the transmission range, so the LoS propagation is predominant. Green’s function is utilized to fully capture the EM characteristics in the LoS propagation. Furthermore, to obtain the tractable channel, the complex channel response between each pair of transmitter/receiver points is investigated. The channel response is composed of the amplitude component and phase component, which display different characteristics in different EM regions \cite{9617121}.

$\bullet$ \emph{Green's Function Based Channel Modeling}

To accurately model the EM characteristics in free space, the homogeneous Helmholtz wave equation is applied, which describes the relationship between the current distribution at the transmitter and the electric field at the receiver \cite{9650519,2021arXiv211108630Z}. Then, Green's function is introduced to solve the Maxwell's equations numerically and to depict the electric field at any receive point of the receiver as \cite{9650519,2021arXiv211108630Z,9765526}. The scalar Green's function without polarization was investigated in \cite{9765526} but a more generalized dyadic Green's function with full polarization was utilized in \cite{9650519,2021arXiv211108630Z}. The dyadic Green's function considers three polarizations at each pair of transmit/receive point. 

$\bullet$ \emph{Complex Channel Response}

For the XL-MIMO over the near-field region, the distances and AoA/AoD between the transmitter and receiver antenna elements vary over the array. To compare clearly the channel responses in different regions, three commonly investigated response models are discussed. Meanwhile, three regions are defined: the radiative near-field region, the Fresnel region, and the far-field region \cite{9617121}. We assume that only one point antenna is considered as the transmitter.
\begin{itemize}
\item \textbf{Radiative near-field region}: The non-uniform spherical wave model is applied where both the amplitude and phase variations over the receiver aperture are noticeable and need to be modeled by the exact propagation distances between the transmitter and particular terminal;
\item \textbf{Fresnel region}: The uniform spherical wave model is investigated, where the amplitude variation can be negligible but noticeable phase variations over the receiver aperture should be considered;
\item \textbf{Far-field region}: The uniform plane wave model is utilized based on the plane wave assumption, where the link distance is much larger than the array dimension. This means that the signal amplitude depends only on the propagation distance between the center of the receiver and the transmitter, and the phase variation is determined by the incident angle.
\end{itemize}

\subsubsection{NLoS Propagation}
The stochastic model is desirable to depict a class of scattered propagation environment even in the near-filed. To model the channel under arbitrary scattering conditions, the Fourier plane-wave based method is utilized based on the fact that every spherical wave can be exactly decomposed into an infinite number of plane waves. Besides, for convenience and tractable analysis, the channel can be presented by the superposition of steering vectors with a particular number of path components.

$\bullet$ \emph{Fourier Plane-Wave Representation}

The incident field and received field can be represented by the respective integral superposition of plane-waves. Then, these two representations are coupled by a scattering kernel integral operator, which links all incident plane-waves and every received plane-wave. Consequently, the channel response can be given by the four-dimensional (4D) Fourier plane-wave representation, which can be decomposed into three terms \cite{9765526}:
\begin{itemize}
\item \textbf{Source response}: maps the excitation current at the particular transmitted point to the certain transmit propagation direction;
\item \textbf{Receive response}: maps the certain receive propagation direction to the induced current at an arbitrary receive point;
\item \textbf{Angular response}: maps each source propagation direction onto each receive propagation direction.
\end{itemize}

Furthermore, to obtain a mathematically tractable channel, the discretized plane waves are defined within the lattice ellipses at the source and receiver, respectively. Then, by adopting the Fourier plane-wave series expansion, the channel response can be approximated by the superposition of discretized plane waves with random Fourier coefficients. More importantly, this channel model holds for both the near-field and far-field regions.

$\bullet$ \emph{Array Response Vector Representation}

The NLoS channel response can be denoted by the superposition of a particular number of array response vectors \cite{2022arXiv220503615L}. The number of array response vectors depends on the number of path components corresponding to the effective scatterers. In the far-field region, the array response vector only relies on the angle of arrival/departure (AoA/AoD) based on the planar wave assumption. However, the array response vector in the near-field region is obtained from the accurate spherical waves, which depends on the AoA/AoD and the distances between the respective transmitter and receiver-scatterer pair.

\subsection{Performance Analysis}
To fully reflect the achievable performance for XL-MIMO systems, two key aspects are discussed in this part: the DoF/EDoF for the XL-MIMO and the electromagnetic information theory for the XL-MIMO.

\subsubsection{DoF for the XL-MIMO}
The DoF, which is the rank of the channel matrix, is always no larger than the minimum between the number of transmitting/receiving antennas. In mMIMO, the available orthogonal communication channels (i.e., the DoF or communication modes) are very limited under the far-field LoS channel since only a single incident angle is available. Based on the spherical-wave assumption, a large range of angles can be achieved by the near-field LoS channel. So exploiting the XL-MIMO can significantly enhance the DoF, even in the LoS propagation.

Based on the prolate spheroidal wave functions, the approximate solution of DoFs for the XL-MIMO is presented \cite{miller2019waves}: for the XL-MIMO with the linear antenna array, DoFs are proportional to the product of the transmitter and receiver length, and inversely proportional to the distance between the transmitter and receiver; for the XL-MIMO with the large surface (CAP and UPA, etc.), the DoFs are proportional to the product of the transmitter and receiver area, and inversely proportional to the square of the distance between the transmitter and receiver.

However, these elegant intuitive approximate results rely on the assumption that the distance between the transmitter and receiver is much larger than their physical sizes and would overestimate the DoF when the link distance is comparable with the physical sizes of transmitter and receiver \cite{9139337}. Based on the 2D sampling theory arguments, the approximate expressions for the DoF are obtained in \cite{9139337}. However, all results above are based on the scalar Green's function model. The EDoF is derived based on the heuristic approach over the EM channel matrix deduced from the dyadic Green's function with full polarizations \cite{9650519}. The EDoF is the equivalent number of SISO systems, which can be approximately calculated by $( {\mathrm{tr}( \mathbf{HH}^H )}/{\| \mathbf{HH}^H \| _{\mathrm{F}}} ) ^2$, where $\mathbf{H}$ is the channel matrix.

\subsubsection{Electromagnetic Information Theory for the XL-MIMO}
The traditional channel capacity theory cannot capture the actual EM characteristics and mismatches the four-dimensional EM fields. In this regard, the electromagnetic information theory is necessary to reveal the fundamental theoretical capacity bound of the EM fields. The EM communication model between two four-dimensional continuous regions is developed in \cite{2021arXiv211108630Z} and the random field instead of the deterministic field is investigated. Then, the mutual information between the parallel linear transmitter and receiver is derived based on the Mercer expansion \cite{2021arXiv211108630Z}. 

\subsection{Signal Processing}
Due to the change in physical sizes and electromagnetic characteristics, the signal processing schemes for the XL-MIMO are different from those for conventional mMIMO. In this subsection, ``channel estimation" and ``beamforming/precoding design" are discussed to track the promising insights for the signal processing in the XL-MIMO.

\subsubsection{Channel estimation}

In this part, we would introduce two channel estimation schemes for the XL-MIMO. It is vital to capture the actual channel features to implement the channel estimation for the XL-MIMO.

$\bullet$ \emph{Reduced-Subspace LS Estimation Scheme}

In \cite{9716880}, the channel of the BS with a UPA is modeled by providing an exact integral expression for the spatial correlation matrix with non-isotropic scattering and directive antennas. The channel spatial correlation matrix is complex to acquire and becomes strongly rank-deficient. Meanwhile, the MMSE-based estimator is difficult to implement due to its high complexity. Indeed, a novel channel estimation scheme called ``reduced-subspace least-squares (RS-LS)" is proposed with the aid of the eigenspace of the spatial correlation matrix, which exploits the array geometry. The RS-LS scheme outperforms the conventional LS estimator and yields similar performance to the MMSE estimator.

$\bullet$ \emph{Polar Domain-Based Estimation Scheme}

To fully capture the near-field channel characteristic, a polar-domain representation, which simultaneously embraces both the angle and distance information, is investigated in \cite{2022arXiv220503615L}. More specifically, the Fresnel approximation is applied to design the polar-domain transform matrix. Then, a specific orthogonal matching pursuit (OMP)-based algorithm as \cite[Algorithm 2]{2022arXiv220503615L} is proposed to obtain the NLoS channel estimates for the XL-MIMO. The proposed estimation scheme is efficient to estimate the channels and provide vital insights for the sparsity representation for the XL-MIMO channels.

\subsubsection{Beamforming/Precoding design}
The beamforming/precoding design determines the system performance for the XL-MIMO. In this part, we introduce processing schemes for the XL-MIMO and discuss the necessity of the beamforming/precoding design in the XL-MIMO.

$\bullet$ \emph{Pattern-Division Multiplexing Technique}

In the XL-MIMO, the EM wave is anticipated to carry the information to approach excellent performance. The XL-MIMO has the ability to generate any electric current density distribution in an anticipatory manner, which allows the system to be designed flexibly for achieving superior performance. Thus the pattern, i.e., the electric current density distribution, is the vital factor to determine the performance. In \cite{2021arXiv211108630Z}, a novel pattern-division multiplexing technique for the CAP-based XL-MIMO system with multiple users is proposed to achieve sum capacity maximization. More specifically, an equivalent transformation for sum capacity maximization is implemented and then the pattern is designed with the aid of iterative optimization. This pattern-division multiplexing technique shows the potential for the XL-MIMO to control the electric current density to achieve excellent performance.

$\bullet$ \emph{Low Computational Precoding Scheme}

With the sharp increase in the number of antennas, the XL-MIMO suffers from beamforming/precoding schemes with extremely high complexity. So it is quite vital to design lightweight signal processing schemes to decrease the computational complexity. In \cite{9779586}, the UPA-based XL-MIMO system with single BS and multiple UEs is investigated based on the channel model in \cite{9765526}. More importantly, a low computational ZF precoding scheme is proposed, where the Neumann series expansion is applied to replace the matrix inversion. The proposed scheme provides important insights into the lightweight processing design for the XL-MIMO.

As discussed above, the signal processing schemes for the XL-MIMO should capture the EM characteristics and have low computational complexity.

\section{Key Challenges \& Solutions}
\begin{figure*}[t]
\setlength{\abovecaptionskip}{-0.1cm}
\centering
\includegraphics[width = 1\textwidth]{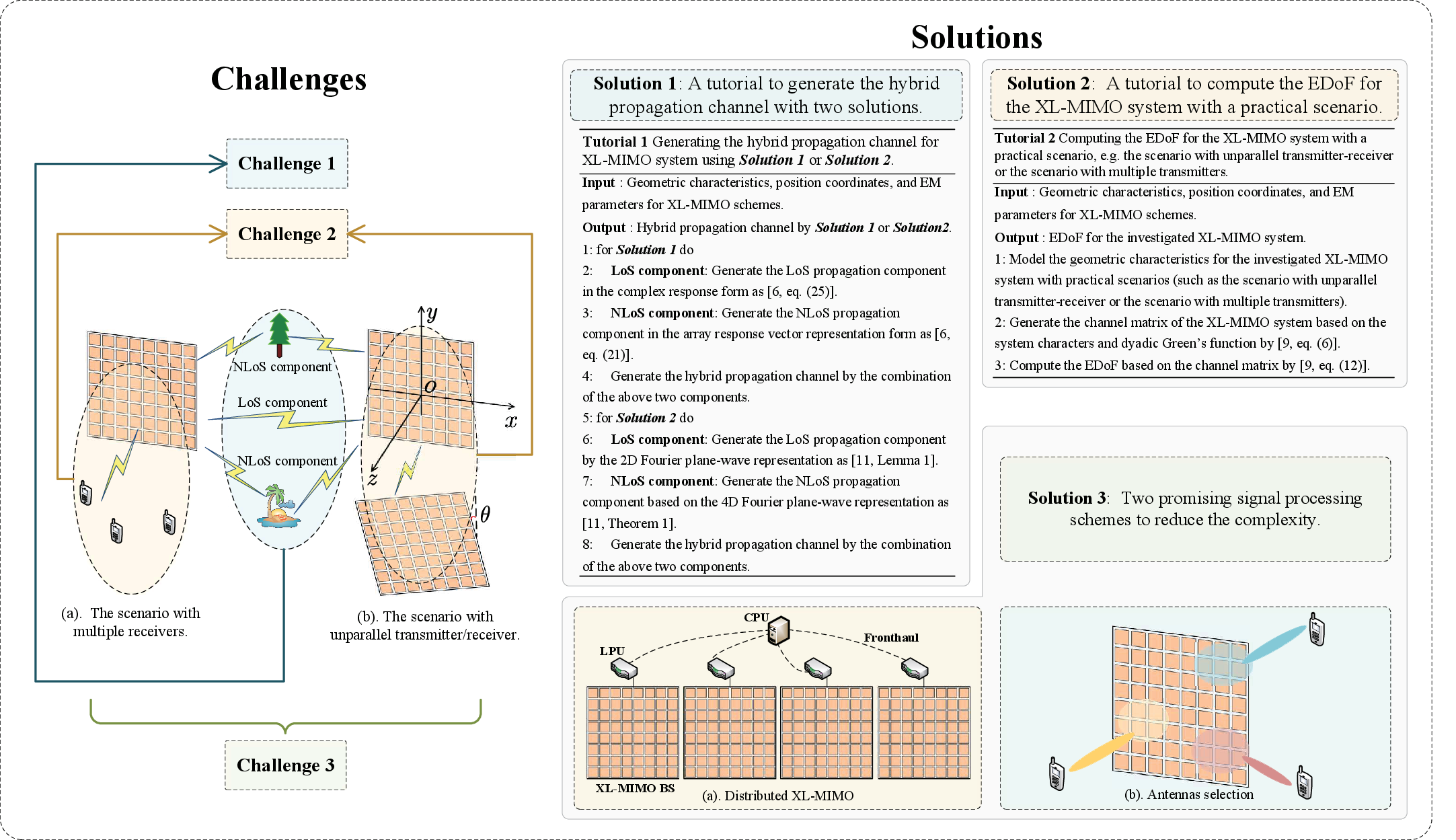}
\caption{Key challenges and respective solutions. \textbf{Challenge 1}: How to model the accurate and tractable channel for the hybrid propagation with the LoS component and the NLoS component? \textbf{Challenge 2}: How to analyze the EDoF performance with practical scenarios (such as unparallel transmitter and receiver as (a) or multiple receivers as (b))? \textbf{Challenge 3}: How to design the processing schemes with acceptable complexity?}
\label{Challenge}
\end{figure*}

In this section, some existing key challenges and respective solutions are discussed. We have summarized these key challenges and respective solutions in Fig. \ref{Challenge}. 

\subsection{Channel Modeling}

The channel modeling for the XL-MIMO has been well summarized in Section~\ref{Sec_Channel} and all channel modeling methods are actual/approximate solutions to Maxwell's equations in specific scenarios. However, in practice, the hybrid propagation channel is probable which embraces both the LoS propagation component and NLoS propagation component. Therefore, accurate and tractable channel modeling for the hybrid propagation channel needs to be further investigated. Some efforts have endeavored to model the hybrid propagation channel \cite{2022arXiv220503615L} based on the array response methods as \cite[eq. (25)]{2022arXiv220503615L}, \cite[eq. (21)]{2022arXiv220503615L} and \cite[eq. (27)]{2022arXiv220503615L}, respectively. Furthermore, another promising solution can be implemented with the aid of Fourier plane-wave representation based on \cite{9765526}. To better clarify the modeling methods, we provide a tutorial as Tutorial 1 in Fig. \ref{Challenge}.


\subsection{Performance Analysis}
Current works on the DoF analysis are based on the over-idealistic scenarios, such as the parallel transmitter and receiver, single transmitter and receiver pair, and scalar Green's function-based channel model, e.t.c. It is necessary to explore the DoF (or EDoF) performance for more practical scenarios. More specifically, the effects of unparallel transmitter and receiver or multiple transmitters/receivers on the DoF (EDoF) performance need to be explored. Meanwhile, a more practical dyadic Green's function-based channel model should also be advocated due to the consideration of full polarization \cite{9650519}.

\begin{figure}[t]
\centering
\includegraphics[scale=0.5]{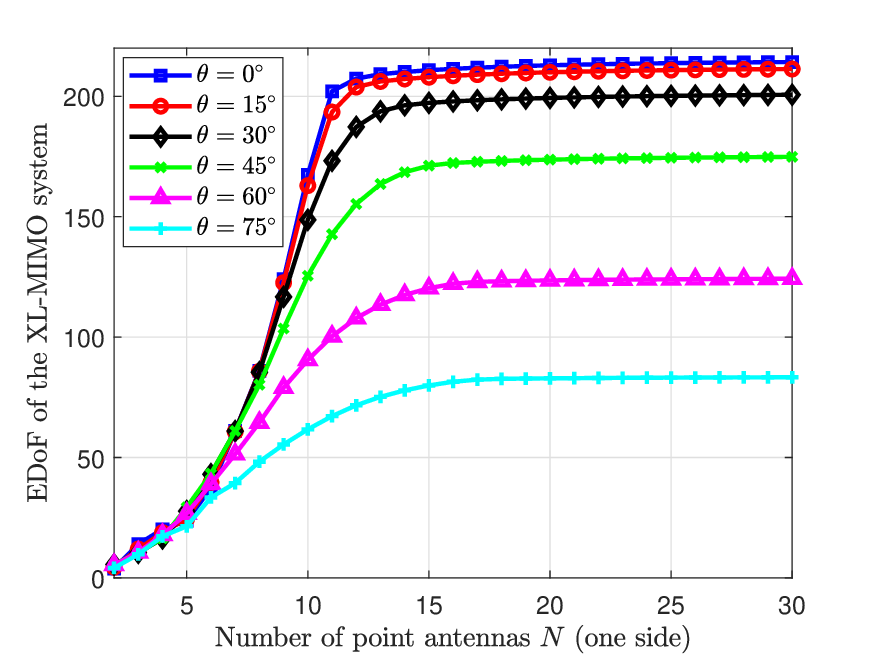}
\caption{EDoF for the system with unparallel XL-MIMO surfaces against $N$. The system consists one transmit surface and one receive surface, where these two surfaces are square with $N\times N$ uniformly distributed point antennas, respectively. The transmitter is assumed to be parallel to the $XY$-plane and the rotation angle between the $XY$-plane and the receiver is $\theta$. The coordinates of the transmitter center and the receiver center are $(0,0,0)$ and $(0,0,7\lambda_0)$, respectively. The side length of the surfaces and the antenna spacing are $L=10\lambda _0$ and ${L}/{N}$, respectively, where $\lambda _0$ is the free-space wavelength.
\label{Unparallel_Surface}}
\end{figure}

\begin{figure}[t]
\centering
\includegraphics[scale=0.5]{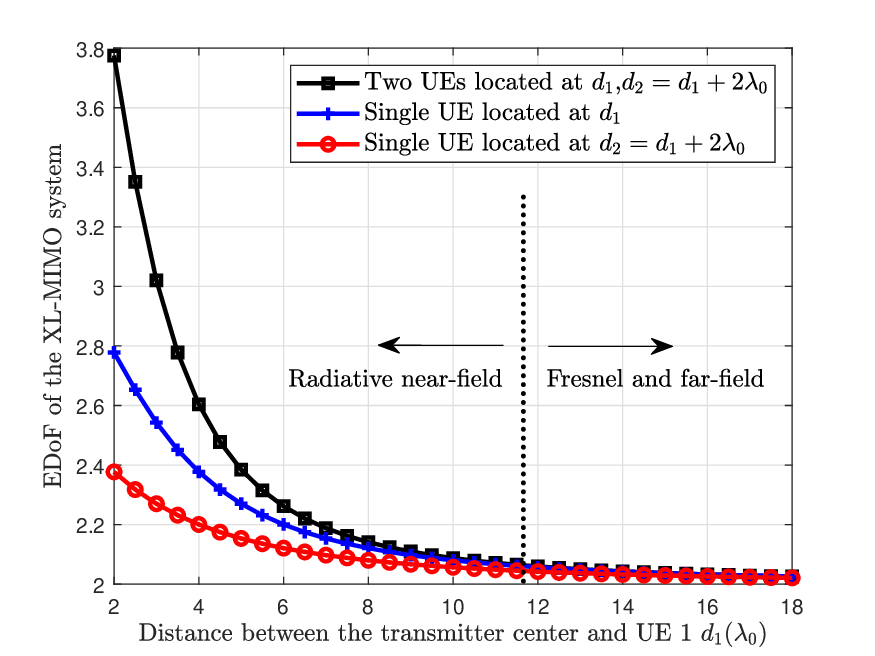}
\caption{EDoF for the XL-MIMO system against $d_1$. The system consists of one transmit surface with $L=5\lambda_0$ side length and $20\times 20$ uniformly distributed point antennas and two/one single-antenna UEs. The transmitter is parallel to the $XY$-plane and the coordinate of the transmitter center is $(0,0,0)$. Two UEs are located at $(0,0,d_1)$ and $(0,0,d_2)$, respectively, with $d_2=d_1+2\lambda_0$. The radiative near-field region, Fresnel region, and far-field region are bounded by $0.62\left({D^3}/{\lambda _0} \right) ^{\frac{1}{2}}=11.66{\lambda _0}$ and ${2D^2}/{\lambda _0}=100{\lambda _0}$, respectively, with $D=\sqrt{2}L$ being the antenna aperture of the BS.
\label{Multi_UEs_Single_Antenna}}
\end{figure}

To analyze the DoF (EDoF) performance with practical scenarios (such as the scenario with unparallel transmitter/receiver and the scenario with multiple receivers), a promising solution is summarized in Tutorial 2 in Fig. \ref{Challenge} (we have ``EDoF" as the performance indicator). Based on Tutorial 2, we analyze two practical scenarios: the scenario with unparallel transmitter-receiver as shown in Fig. \ref{Unparallel_Surface} and the scenario with multiple transmitters as shown in Fig. \ref{Multi_UEs_Single_Antenna}. We consider the UPA-based XL-MIMO with point antennas, and, as for the other parameters set, please refer to \cite{9650519}. As shown in Fig. \ref{Unparallel_Surface}, we observe that the scenario with parallel transmitter and receiver achieves the highest EDoF performance. Besides, with $N$ increasing and the antenna spacing decreasing, the EDoF would increase fast at first and then reach a maximum value. As observed in Fig. \ref{Multi_UEs_Single_Antenna}, the EDoF for the scenario with two UEs is lower than the addition of the EDoFs with only UE $1$ and only UE $2$ since the coupling effect caused by these two UEs may have a great effect on the EDoF performance. Moreover, when $d_1$ increases, the EDoF with two UEs approaches to that of a single UE scenario due to the limited EM characteristics in the  Fresnel region and far-field region.


\subsection{Signal Processing}

The signal processing schemes with high complexities would be implemented in the XL-MIMO system due to the large array size. Therefore, it is challenging to design processing schemes with acceptable complexity. Some promising solutions are discussed as follows.

$\bullet$ \emph{Distributed XL-MIMO Implementation}

Inspired by the promising cell-free massive MIMO technology, the distributed XL-MIMO implementation is advocated as ``Solution 3 (a)" in Fig. \ref{Challenge} to reduce the computation overhead of BSs. The distributed XL-MIMO implementation consists of XL-MIMO BSs with local processing units (LPUs), central processing units (CPUs) with high computation processing ability, and fronthaul links. With the aid of distributed topology and assistance from the CPU, different processing schemes can be implemented flexibly according to different computation requirements \cite{9113273}.

$\bullet$ \emph{Antenna Selection and Antenna Activation}

In practical communication, we may not need to simultaneously use all antennas in the XL-MIMO array. In some certain scenarios, only a partial group of antennas need to be activated to serve a particular UE as ``Solution 3 (b)" in Fig. \ref{Challenge}. Some optimization problems can be established to select a particular group of antennas to serve each UE or turn on/off particular antennas to satisfy different design requirements.

The high processing complexity would correspondingly also lead to the high signal processing latency. To further lower the processing complexity and the processing latency, deep learning (DL)-based processing schemes, e.g. DL-based beam training, and low-complexity-based processing schemes \cite{9716880}, \cite{9779586}, are advocated. Besides, the semantic-based scheme is applicable and promising since it can reduce the transmitted data size by extracting semantic information from the original source data \cite{yang2022semantic}.

Moreover, the power consumption and energy efficiency issues are also vital challenges for the XL-MIMO. To reduce the power consumption, one possible method is to deploy the XL-MIMO with a few RF chains, where each RF chain is connected to a certain module of antennas. Moreover, the hybrid-beamforming and mixed-resolution ADCs architecture are also potential solutions. Furthermore, the antenna selection, the modular architecture and spare processing based schemes need to be newly designed to achieve high energy efficiency.


\section{Future Directions}

\subsection{Distributed XL-MIMO Implementation}
As discussed in Section III, the distributed XL-MIMO implementation is promising for its uniform coverage and outstanding distributed processing ability. However, the distributed XL-MIMO still needs to be investigated. With the versatile topology of distributed XL-MIMO, different signal processing schemes can be implemented distinguishing from the performance requirement and complexity trade-off. Meanwhile, the BS and antenna selections are also vital for the practical distributed implementation. Besides, the hardware cost and synchronization issues are indispensable for the XL-MIMO. The low-resolution ADC and hybrid-beamforming are advocated to lower the hardware cost, and the reference signal design for the over-the-air reciprocity calibration is an efficient method to address the synchronization issue. More importantly, the distributed optimization at each XL-MIMO BS is interesting, and distributed learning is regarded as a possible solution.

\subsection{Semantic Communications}
Further performance breakthroughs of the XL-MIMO system can be achieved by using novel communication paradigms, i.e., semantic communications (SemCom) \cite{yang2022semantic}. Through extracting and transmitting task-related semantic information from source data, SemCom can improve network efficiency and reduce the wireless data transmission burden. A joint semantic channel encoding scheme can be designed in the SemCom-aided XL-MIMO system. Specifically, a semantic extraction module can learn semantic features from the source data. The learned features have a smaller data size than the original data, which reduces the wireless transmitting time and thus helps to implement URLLC. Furthermore, the XL-MIMO technique can enhance the SemCom system. Specifically, an attention feature module in the semantic encoder can use the learned features to produce a sequence of scaling parameters. These parameters represent the importance of semantic features and can guide resource allocation and beamforming design. For example, in the case of limited network resources, finer beamforming schemes can be used when transmitting relatively more important semantic information to ensure error-free communication.

\subsection{Integrated Sensing and Communications}
Since wireless signals and various sensors are ubiquitous in our daily life, a new paradigm shift in wireless network design could be integrated sensing and communications (ISAC).
By integrating sensing and communication functionalities, ISAC can achieve efficient usage of wireless resources. By considering the ISAC design in XL-MIMO, two functionalities can mutually benefit each other. Specifically, the sensory data can be collected and utilized to assist the beamforming algorithm design to enhance communication performance. More importantly, the localization and tracking can significantly benefit from the abundant DoFs \cite{2022arXiv220313035Z}. Nonetheless, how to exploit the DoF offered by the XL-MIMO for the localization and tracking is a promising future direction.
\subsection{Wireless power transfer}
Wireless power transfer (WPT) is regarded as a promising technology to address the critical energy supply issues in the future wireless networks. By investigating XL-MIMO, the WPT can be implemented in the near-field region, which is the emerging topic ``near-field WPT" \cite{2022arXiv220313035Z}. Due to the efficient transfer with little energy pollution, the near-filed WPT protocol and the relationship between the DoF offered by XL-MIMO and the WPT capabilities can be important future directions.

\vspace{-0.7cm}

\section{Conclusions}
In this article, we investigated the fundamentals, challenges, solutions, and future research directions for the promising XL-MIMO technology. We introduced four general XL-MIMO schemes and discuss their characteristics/relationships. The fundamentals of  ``channel modeling", ``performance analysis" and ``signal processing" aspects for the XL-MIMO were comprehensively reviewed. Then, some challenges for the above three aspects were discussed and respective solutions were provided. More importantly, we proposed two tutorials for the hybrid propagation channel modeling and the EDoF computations for practical scenarios. Numerical results investigated the EDoF performance for the scenario with unparallel XL-MIMO surfaces or the scenario with multiple UEs. Finally, we provided promising directions and insightful inspiration for future research on XL-MIMO.

\bibliographystyle{IEEEtran}
\bibliography{IEEEabrv,Ref}
\end{document}